\documentclass[11pt]{article}
\usepackage{moriond}

\def\be{\begin{equation}}
\def\ee{\end{equation}}
\def\bea{\begin{eqnarray}}
\def\eea{\end{eqnarray}}

\begin{document}
\vspace*{4cm}
\title{TeV Dark Matter detection by Atmospheric \v{C}erenkov Telescopes \footnote{Contribution to the proceedings of the 2005 Moriond workshop ``Very High Energy Phenomena in the Universe''. Based on~\cite{Hooper:2004vp}.}}

\author{Francesc Ferrer}

\address{CERCA, Department of Physics, Case Western Reserve University,
10900 Euclid Avenue, Cleveland, OH 44106-7079, USA}

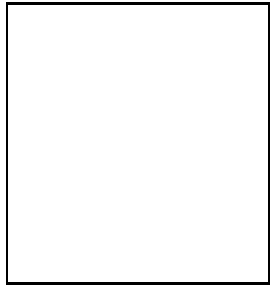
\maketitle\abstracts{Ground based Atmospheric \v{C}erenkov Telescopes have recently unveiled a TeV gamma-ray signal from the direction of the Galactic Centre. We examine whether these $\gamma$-rays, observed by the VERITAS, CANGAROO-II and HESS collaborations, 
may arise from annihilations of dark matter particles. Emission from nearby dwarf spheroidals, such as Sagittarius, could provide a test of this scenario.  
}

\section{Introduction}
Detections of TeV $\gamma$-rays from the Galactic Centre~(GC) region have been made by Atmospheric \v{C}erenkov Telescopes~(ACTs). The VERITAS\cite{Kosack:2004ri} collaboration observed the central region of the Galaxy with the Whipple telescope at high zenith angle, resulting in a high energy threshold of $\sim 2.8$ TeV. A $3.7\sigma$ signal with an integral flux of $F_\gamma (> 2.8\ \rm{TeV}) \sim 1.6 \times 10^{-8}\ \rm{m}^{-2}\ \rm{s}^{-1}$ was reported. For ACTs located in the southern hemisphere, the energy threshold towards the GC is an order of magnitude lower. CANGAROO-II\cite{Tsuchiya:2004wv} claimed a $\sim 10\sigma$ detection above $\sim 250$ GeV with an integrated flux of $F_\gamma (> 250\ \rm{GeV}) \simeq 2 \times 10^{-6}\, \rm{m}^{-2}\ \rm{s}^{-1}$ and a very soft spectrum, $\propto E^{-4.6}$. The HESS\cite{Aharonian:2004wa} collaboration detects\footnote{Continued observations by HESS have confirmed this result with an increased significance of $34\sigma$~\cite{rolland}.} a  $\sim 9\sigma$ excess with a much harder spectrum, $\propto E^{-2.2}$. This rather hard spectrum is obviously not consistent with the steeper profile obtained by CANGAROO-II. The large flux at low energies implied by the CANGAROO result would have been detected by HESS in a matter of minutes. On the other hand, the flux seen by Whipple is roughly consistent with HESS. Further observations in the coming years will help in clarifying the situation.

Although the origin of this emission is unknown, there are several possible, though unlikely, astrophysical sources in the field of view. Alternatively, the possibility that these TeV $\gamma$-rays may be due to annihilating dark matter particles was put forward in~\cite{Hooper:2004vp}, and further studied in~\cite{rolland,Horns:2004bk}.

\section{Annihilating dark matter}
The flux of $\gamma$-rays observed from the GC, if due to dark matter annihilation, is given by:
\begin{equation}
\frac{{\rm d}\Phi_\gamma (\psi, E_\gamma)}{{\rm d}E_\gamma}
	 = \langle\sigma v\rangle
                               \frac{\rm{d}N_\gamma}{\rm{d} E_\gamma}
                               \frac{1}{4\pi M_X^2}\int_{\rm los}
                               \rm{d}l(\psi)\ \rho^2(r).
\label{flux}
\end{equation}
Here, $\psi$ is the angle between the line-of-sight (los) and the
GC, $\langle\sigma v\rangle$ is the dark matter
annihilation cross-section averaged over its velocity distribution,
$\rm{d}N_\gamma/\rm{d} E_\gamma$ encodes the $\gamma$-ray spectrum 
and $\rho(r)$ is the dark matter density at a distance $r$ from the
GC. 

Annihilations of dark matter particles can produce $\gamma$-rays
in several ways (see e.g.~\cite{Bergstrom:1997fj}). A continuum of $\gamma$-rays results from the hadronization and decay of $\pi^0$'s generated in the cascading of annihilation products. Also, monoenergetic $\gamma$-ray lines are
produced as dark matter particles annihilate via the modes $X X
\rightarrow \gamma \gamma$ and $X X \rightarrow \gamma Z$. However, these
processes are loop suppressed
and thus yield much smaller fluxes than continuum emission.
A good fit to the continuum spectrum obtained by fragmentation Monte Carlo codes is provided by\cite{Hooper:2004vp}:
\be
\frac{\rm{d}N_\gamma}{\rm{d} E_\gamma}  \simeq \frac{0.73}{M_X}\frac{ {\rm e}^{-7.76 E_\gamma/M_X}}{(E_\gamma/M_X)^{1.5} + 0.00014} \equiv 0.73 \, {\rm TeV}^{-1}\left(M_{X,TeV}\right)^{-1}  \left. {\cal F}(x) \right|_{x=E_{\gamma}/M_X},
\ee
where $M_{X,TeV}$ is the dark matter particle mass in units of 1 TeV.

The data from CANGAROO-II appears to fit the spectrum reasonably well for a 1--3 TeV dark matter particle\cite{Hooper:2004vp}. However, this is in conflict with the Whipple and HESS observations. The spectral shape observed by HESS requires\cite{Horns:2004bk} a very heavy particle ($\sim 10$ TeV). Moreover, if dark matter annihilation indeed produces the TeV $\gamma$-rays observed by ACTs, a lower energy component is expected to which
EGRET (or in the future GLAST) is, in principle, sensitive. The upper limit placed by EGRET\cite{Hooper:2002ru} excludes any particle lighter than about 3.5--4 TeV, for a spectrum normalised to Whipple. From now on, we will assume that continuum emission from a heavy particle ($\sim 10$ TeV) is contributing to the observed TeV $\gamma$-rays.

We now turn our attention to the annihilation cross-section, which also determines the dark matter relic thermal abundance. In the simplest situation the relic abundance is approximately given by $ \Omega_X h^2 \sim (\langle\sigma
v\rangle/3\times10^{-27}\,\rm{cm}^{3}\rm{s}^{-1})^{-1}$ . Although there are certainly exceptions to
this estimate, it is reasonable to consider $\sigma
v_{26}\sim 1$, where we expressed $\langle\sigma
v\rangle$ in units of $3 \times 10^{-26} \,
\rm{cm}^3 \, \rm{s}^{-1}$, as an upper limit for a thermal relic which makes up the cold
dark matter. Of course, a non-thermally produced dark matter particle could have a larger $\sigma v_{26}$.

Finally, we need to consider the dependence of Eq.~\ref{flux} on the dark matter distribution. This part is usually parametrised by
$ J (\psi) = (8.5 \rm{kpc})^{-1} (0.3 \rm{GeV}/\rm{cm}^3)^{-2} \int_{\rm los} \rm{d}l(\psi)\ \rho^2(l)$. Defining $\overline{J(\Delta \Omega)}$ as the average of
$J(\psi)$ over the solid angle $\Delta \Omega$ ($\sim
5\times10^{-5}$ sr for ACTs), we obtain:
\begin{equation}
\frac{\rm{d} \Phi_\gamma}{\rm{d}E_\gamma}  (\psi, E_\gamma) \simeq 4 \times 10^{-12}\
 \rm{cm}^{-2}\rm{s}^{-1} \rm{TeV}^{-1}
{\cal F}\left(E_\gamma/M_X\right)
 \sigma v_{26}
 \left(M_{X,\rm{TeV}}\right)^{-3} \overline{J (\Delta\Omega)}
 \Delta\Omega.
\label{fluxnum}
\end{equation}
The value of $\overline{J(\Delta\Omega)}$ can vary
a great deal depending on the assumed dark matter distribution,
e.g. for an NFW halo profile\cite{nfw} $\overline{J(5 \times
10^{-5}\, \rm{sr})} \equiv \overline{J_{-5}}\simeq 5.6 \times 10^3$, while a Moore {\it et
al.} halo profile\cite{moore} yields a considerably larger value,
$\overline{J_{-5}} \simeq 1.9 \times 10^6$.

It is possible, however, that these highly cusped profiles deduced
from N-body simulations, do not accurately represent the
distribution of dark matter in our halo. Simulations cannot resolve halo profiles on scales smaller than
roughly 1 kpc, and must rely on extrapolations in the innermost
regions of the GC\cite{Binney:2003sn}. Also, these
simulations model halos without baryons. Baryons,
that had been significantly heated in the past, may
have expanded outward, gravitationally pulling the dark matter and
thus reducing its density in the innermost regions. Conversely, as
baryonic matter loses energy through radiative processes, it will fall
deeper into the Galaxy's gravitational well, pulling dark matter along
with it. Alternatively, adiabatic accretion of
dark matter onto the Super-Massive Black Hole (SMBH) at the GC may have produced a density `spike' in the halo profile\cite{spike}. Although, such a spike
could have been destroyed in a series of hierarchical mergers or stellar encounters\cite{merrit}. 

Astrophysical observations constrain the dark
matter distribution in the GC. The observed number of microlensing events in
the direction of the Bulge can be
used\cite{be}  to exclude cuspy halo profiles with inner power-law indices
greater than about 0.3 (NFW and Moore {\it et al.}
profiles have indices of 1.0 and 1.5, respectively). Also, if dark matter dominates the GC, dynamical friction would cause the central bar, which is known to be fastly rotating, to decelerate on a few bar rotation time scales\cite{bat}.  The claim in~\cite{kzs} that a reasonable agreement between observational data and cuspy halo profiles can be reached including adiabatic compression is, in large part, due to the adoption  of a
significantly  lower microlensing optical depth towards the GC.
Given this wide range of opinions regarding
the halo dark matter profile, it may be prudent not to exclude any of
these models from our discussion.

We are ready to discuss the annihilation cross-sections and
values of $\overline{J_{-5}}$ needed to
accommodate the observations (see Fig. 4 in~\cite{Hooper:2004vp}). 
For an NFW halo profile, very large cross-sections, $\sigma v_{26} \sim 10^3$ would be required to match the fluxes
detected, excluding a thermal relic. On the
other hand, for a more centrally concentrated halo, cross-sections on the order  of $\sigma v_{26} \sim 1$ could suffice to
match the observed fluxes. Spiked density profiles could readily
accommodate the observed flux.  Indeed, one could even
dispense with the need for the spike to surround the central SMBH, 
provided that another pregalactic  VMBH  which had retained its spike 
would be within a few parsecs of the GC. In this sense, it is intriguing to note the appearance of a bump in the HESS data at an angle $\psi \sim 0.17^\circ$ corresponding to a distance of $\sim 20pc$ from the GC\cite{Aharonian:2004wa}.
Let us also mention, that the good agreement with a point source  favours a centrally concentrated profile for the superior angular resolution of HESS data\cite{Horns:2004bk}.

\section{Emission from Sagittarius}

Dwarf spheroidal galaxies (dSphs) warrant attention because they are amongst the most extreme dark matter dominated environments. Thus, large dark matter annihilation rates and related $\gamma$-ray fluxes are predicted from these regions\cite{Evans:2003sc,Hooper:2003sh}. Unlike the GC, dSphs are dark matter dominated and do not contain substantial amounts of gas or stars. Therefore, observation of TeV emission from one or more dSphs would provide strong evidence for annihilating dark matter. 

Moreover, the astrophysical observations constraining the amount of dark matter in the GC do not apply to dSphs. Two sets of models of dSphs were developed in~\cite{Evans:2003sc}. The closest confirmed dSph is Sagittarius, located $\sim 24$ kpc away from the Sun. Assuming a Moore profile for Sagittarius, $\overline{J_{-5}} \sim 3 \times 10^4$, which using Eq.~\ref{fluxnum}, for a 10 TeV particle with $\sigma v_{26} \sim 1$, gives:
\begin{equation}
\frac{{\rm d} \Phi_\gamma}{{\rm d} E_\gamma} (\rm{Sag}, 1 \rm{TeV}) \simeq 1.8 \times 10^{-9}
 \rm{cm}^{-2}\rm{s}^{-1} \rm{TeV}^{-1}.
\end{equation}

This corresponds to 10\% of the flux recorded by HESS telescope from the GC at 1 TeV\cite{Aharonian:2004wa}, well in reach of its sensitivity. Thus, once sufficient exposure in the direction of Sagittarius is attained, TeV emission from annihilating dark matter particles in Sagittarius could be discovered. 

It is also possible that annihilating dark matter causes only part of the signal from the GC. Some astrophysical source would then give most of the signal. In this respect, the HESS collaboration has reported the discovery of eight point sources in the Galactic Plane\cite{Aharonian:2005jn}. Since these sources are not, in principle, located in a dark matter dominated region, they are probably of astrophysical origin. However, in the Sagittarius dSph we expect the signal from dark matter annihilations to be enhanced, whereas astrophysical sources masking the signal are not expected to be present in this dark matter dominated environment.

\section{Discussion}

We have studied the possibility that annihilating dark matter in the GC produces the flux of $\gamma$-rays observed by ACTs. If this is the mechanism behind the observed emission, the particles making the dark matter should have a sizable annihilation cross-section, $\sigma v_{26} \sim 1$, and follow a cuspy distribution in the GC.

To accommodate the observations, Whipple and HESS in particular, the dark matter candidate must be a few TeV or heavier. In models of softly broken supersymmetry, the neutralino is not usually expected to be so heavy, being more natural near the electroweak scale. However, there are other viable dark matter candidates in addition to the neutralino. These include stable particles from the messenger sector of gauge mediated supersymmetry breaking models or Kaluza-Klein dark matter in models with a universal extra dimension\cite{Hooper:2004fh}.

If dark matter is at the root of the TeV radiation from the GC, then observations in the direction of the Sagittarius dSph should reveal a TeV $\gamma$-ray signal. The absence of such a signal would suggest that the observed emission is not related to dark matter annihilations, while a positive detection would support the possibility that ACTs have observed TeV dark matter.

\section*{Acknowledgments}
Part of this work was done in collaboration with D.~Hooper, I.~de la Calle Perez, J.~Silk and S.~Sarkar. The author is supported by the U.S. Department of Energy and NASA at Case.

\end{document}